\documentclass[fleqn,twoside,twocolumn,nofootinbib,showkeys]{revtex4} 
\usepackage[nocpr]{ujp} 

\begin{document}
\title[Polarization Dependences of Terahertz Radiation]
{POLARIZATION DEPENDENCES OF TERAHERTZ RADIATION EMITTED BY HOT
CHARGE\\
CARRIERS IN \boldmath$p$-Te}%
\author{P.M.~Tomchuk}
\affiliation{Institute of Physics,
Nat. Acad.
of Sci. of Ukraine}
\address{46, Nauky Ave., Kyiv 03680, Ukraine}
\email{ptomchuk@iop.kiev.ua}
\author{V.M.~Bondar}%
\affiliation{Institute of Physics,
Nat. Acad.
of Sci. of Ukraine}%
\address{46, Nauky Ave., Kyiv 03680, Ukraine}%
\author{L.S.~Solonchuk}
\affiliation{Institute of Physics,
Nat. Acad.
of Sci. of Ukraine}%
\address{46, Nauky Ave., Kyiv 03680, Ukraine}%
\email{ptomchuk@iop.kiev.ua}%
\udk{533.9} \pacs{42.72. Ai, 72.20.Ht,\\[-3pt] 61.72.uf, 72.20}
\razd{\secix}

\autorcol{P.M.\hspace*{0.7mm}Tomchuk, V.M.\hspace*{0.7mm}Bondar,
L.S.\hspace*{0.7mm}Solonchuk}

\setcounter{page}{135}%

\begin{abstract}
Polarization dependences of the terahertz radiation emitted by hot
charge carriers in $p$-Te have been studied both theoretically and
experimentally. The angular dependences of the spontaneous radiation
emission by hot carriers is shown to originate from the anisotropy
of their dispersion law and the anisotropy of the dielectric
permittivity of a tellurium crystal. We have shown that the
polarization dependences of radiation are determined by the angle
between the crystallographic axis $C_{3}$ in $p$-Te and the
polarization vector; they are found to have a periodic character.
\end{abstract}
 \maketitle

\section{Introduction}

Free charge carriers can neither absorb nor emit light, because the
laws of energy and momentum conservation cannot be satisfied
simultaneously in such processes. These processes become possible if
\textquotedblleft a third body\textquotedblright\ participates in
them. Various impurities, lattice vibrations, or a boundary can play
this role. Which phenomenon (absorption or radiation emission)
dominates at that depends on external \mbox{conditions.}

If charge carriers are in the thermodynamic equilibrium state, and if the
semiconductor is irradiated with an external electromagnetic flux, the
processes of light absorption by free carriers dominate. On the other hand,
if no external radiation is present, and if carriers are heated up by an electric
field applied to the semiconductor, then the processes of light emission by
free carriers prevail. This radiation emission belongs mainly to the
terahertz frequency range.

In semiconductors, the dispersion law for charge carriers and the
mechanisms of their scattering are anisotropic, so that the
spontaneous emission by hot carriers depends on the polarization.
Earlier, we have studied similar polarization dependences with the
use of $n$-Ge as an example \cite{1,2,3}. However, despite the fact
that the dispersion law of electrons in the minima (valleys) of the
Brillouin zone in multivalley semiconductors of the $n$-Ge type has
a pronounced anisotropic character, the valleys themselves are
arranged symmetrically in this zone. As a result, the angular
dependences of the spontaneous radiation emission by hot electrons
arise only at certain orientations of a heating electric field.
Namely, in crystals with the cubic symmetry of the $n$-Ge and $n$-Si
type, the polarization dependences of the spontaneous radiation
emission by hot electrons appear only at such orientations of a
heating electric field, at which the temperatures of electrons in
different valleys is also different. One can also break the cubic
symmetry and, hence, induce the polarization dependences by applying
a unidirectional pressure to a  \mbox{semiconductor \cite{1}.}

An absolutely different situation takes place in $p$-Te, to which this
work is devoted. Tellurium is a considerably anisotropic
semiconductor, although, provided that certain restrictions are
imposed upon the hole concentration \cite{4}, the isoenergetic hole
surfaces in tellurium can be accepted with a sufficient accuracy in
the form of ellipsoids of revolution, similarly to what takes place
for electrons in $n$-Ge and $n$-Si. However, there is a basic
difference. Namely, the rotation axes of ellipsoids for the electron mass tensor in
$p$-Te (there are six ellipsoids) are parallel to one another and to
the axis $C_{3}$. At the same time, in $n$-Ge and $n$-Si, the
ellipsoids of the electron mass tensor are oriented differently, but so that the
symmetry of a crystal as a whole remains cubic. Therefore, the
polarization dependences caused by the presence of the mass tensor of charge
carriers can be observed
in tellurium in their bare form, i.e. without any modification
related to different contributions of different valleys to the radiation
process, as it takes place in $n$-Ge and $n$-Si.

The processes of radiation absorption and emission by free carriers
can be studied in the framework of various mathematical methods. In
works \cite{1,6}, we suggested a method to study such processes,
which has definite advantages in comparison with the available
methods. Those advantages consist in that a common approach can be
used to analyze absorption and emission of light by free carriers,
in both the classical (when the light quantum energy, $\hbar \omega
$, is much lower than the thermal energy of carriers, $kT_{e}$) and
quantum-mechanical (when $\hbar \omega >kT_{e}$) frequency ranges.
In this approach, the results can be obtained in the analytical form
even if the anisotropy of both the dispersion law and the scattering
mechanisms is taken into account. Just this method was used in
\mbox{this work.}

\section{Theory and Adopted Model}

In accordance with works \cite{5,7}, the dispersion law for holes in $p$-Te
is taken in the form
\begin{equation}
\varepsilon({\bf p})=\frac{p_{\bot}^{2}}{2m_{\bot}}+\frac{p_{\parallel}^{2}}{%
2m_{\parallel}},   \label{eq1}
\end{equation}
where $p_{\parallel}$ and $p_{\bot}=(p_{x}^{2}+p_{y}^{2})$ are the
longitudinal and transverse, respectively, components of the
momentum vector ${\bf p}$ with respect to the direction of the axis
$C_{3}$ in the crystal. For $p$-Te, according to work \cite{8}, we
have $m_{\parallel}=0.26\,m_{0}$ and $m_{\bot}=0.11\,m_{0}$, where
$m_{0}$ is the free electron mass.

Let an external constant electric field be applied to a $p$-Te
specimen, and let this field heat up holes. On the basis of our
experimental conditions (the lattice temperature $T=4~$K and the
concentration of ionized impurities $n_{i}\approx 5\times
10^{14}~\mathrm{cm}^{-3}$), we may suppose \cite{9,10} that the
relaxation of hole momenta is driven by the hole scattering at ionized
impurities, and the relaxation of the hole energy is determined by the inelastic hole
scattering at acoustic and optical lattice vibrations. In the
framework of this model, if the temperature is low, the
contribution to the energy exchange between holes and the lattice
can be given by the interaction between holes and only those optical
vibrations of the lattice, whose Debye temperature is the
lowest. In particular, for $p$-Te, it is the $A_{2}$ optical mode,
for which the Debye temperature equals 138~K \cite{11}.

Below, we describe a scheme of the construction of the theory of spontaneous
radiation emission by hot carriers in brief, readdressing the reader to work
\cite{6} for details.

Hence, in order to obtain the collision integral that involves the
influence of the electromagnetic wave field on the scattering processes of free
carriers, let us do it as follows. While deriving the collision integral,
we use the wave
functions of free carriers in the electromagnetic wave field
\[ \Psi _{ {\bf p}} =\frac{1}{V}\exp \,\left(\! {\frac{i}{\hbar }\, {\bf
p}\, {\bf r}} \!\right)\times\]\vspace{-5mm}
\begin{equation}
\label{eq2} \exp \left\{\! {-\frac{i}{\hbar }\int\limits_0^t
{d{t}'\sum\limits_j^3 {\frac{1}{2m_j }\left( \!{p_j
-\frac{e}{c}\,A_j ({t}')}\! \right)^2} } }\! \right\}\!,
\end{equation}
instead of their basic wave functions. Here, $V$ is the volume,
$t^{\prime }$ the time, $e$ the carrier charge, $c$ the velocity of
light, and ${\bf A}(t)$ the vector-potential in the dipole
approximation, ${\bf A}(t)={\bf A}^{(0)}\cos \,\omega t.$ Note that
the inclusion of the electromagnetic wave field into the collision
integral, rather than into the left-hand side of the kinetic
equation as an external force, is valid if the inequality $\omega
\tau >1$, where $\tau $ is the relaxation time, is satisfied.

As was shown in work \cite{6}, the collision integral for the scattering
of carriers by ionized impurities can be obtained with the use of basis (\ref{eq2}) in
the form
\[
\Hat{I} f=4\,e^4\,N_i\times\]\vspace{-5mm}
\[\times\!\!
\sum_{l=-\infty }^\infty\! {\int \!\!d{\bf p}'\!\frac{f({\bf
p}')\!-\!f({\bf p})}{\left\{\! \chi _\bot \!({\bf p}_\bot
\!\!-\!{\bf p}'_\bot )^2\!\!+\!\chi _{\parallel } (p_{\parallel}\!
-\!{p}'_{\parallel } )^2\!\!+\!\!(\hbar / {r_D^{(0)} })^{2}
\!\right\}^{\!2}}\times }
\]\vspace{-5mm}
\begin{equation}
\times  \Im _{l}^{2} \left(\! \frac{e}{c\,\hbar \,\omega
}\sum_{j=1}^3 A_j^{(0)} \frac{p_j -{p}^{\prime}_j }{m_j }
\!\right)\delta (\varepsilon \,({\bf p})-\varepsilon ({\bf
{p}'})-l\,\hbar \,\omega ),
\end{equation}
Here, $f({{\bf p}})$ is the distribution function of charge carriers
over their momenta ${\bf p}$'s, $N_{i}$ the concentration of ionized
impurities, and $\Im _{l}(x)$ the Bessel function of the $l$-th
order. Moreover, we
have additionally considered the tensor character of the dielectric
permittivity of tellurium in formula (3) in contrast to work \cite{6}. In particular, in the
coordinate system with the axis $0z$ directed along the axis $C_{3}$
in tellurium, we have $\chi _{xx}=\chi _{yy}\equiv \chi _{\bot }$
and $\chi _{zz}\equiv \chi _{\parallel }$. Expression (3) includes
low-frequency values of $\chi _{ii}$. The parameter $r_{D}^{(0)}$ is
the radius of charge screening by free carriers. In particular, in
the case of non-degenerate statistics,
\begin{equation} \left(\! {\frac{1}{r_{D}}}\!\right) ^{\!2}=\frac{4\pi
\,e^{2}n}{kT_{p}}, \label{eq3}
\end{equation}%
where $n$ is the concentration of free carriers, and $T_{p}$ is their
temperature.

If we are not interested in a special action of powerful laser
pulses, the argument of the Bessel function in Eq.~(3) is, as a rule,
less than 1. Therefore, the multiplier $\Im _{l}^{2}(\ldots )$
in formula (3) can be expanded in a series, and only the first term
of the series can be retained. In addition, in what follows, we will
confine the consideration to one-quantum processes, i.e. only the
terms in expression (3) corresponding to $l=\pm 1$ will be taken
into account. Then, by multiplying expression (3) by $\varepsilon
({\bf p})$ and integrating the product over $d{\bf p}$, we obtain
the variation of the electron system energy per unit time, which is
related to the processes of absorption and emission of light quanta, $\hbar
\omega $:
\[
p=\int {d{\bf p}\,\hat {I}f=P(+)+P(-)},
\]\vspace*{-5mm}
\[ P(\pm )=\pm \frac{e^6}{c^2\,\hbar \,\omega }\times\]
\[ \times\int \frac{d{\bf p}\,d{\bar {p}}'\,f({\bf p}')\,\delta
\{ \varepsilon ({\bf p})-\varepsilon ({\bf p}')\pm \hbar \omega
\}}{\{ \chi _\bot ({\bf p}_\bot -{\bf p}'_\bot )^2+\chi _{\parallel
} (p_{\parallel } -{p}'_{\parallel } )^2+(\hbar  / {r_D })^2
\}^{2}}\times\]\vspace*{-5mm}
\begin{equation}
\label{eq4} \times\left(\! \sum A_j^{(0)} \frac{p_j -{p}'_j }{m_j }
\!\right)^{\!2}\! .
\end{equation}
While calculating integral (\ref{eq4}), let the Maxwell function with the
effective hole temperature $\theta _{p}\equiv kT_{p}$ be used as a
distribution function for hot holes,
\begin{equation}
\label{eq5} f({\bf p})=\frac{n}{(2\pi \theta _p )^{3/2}\;m_\bot
\sqrt {m_{\parallel } } }\exp ( -\varepsilon ({\bf p})/ {\theta _p
}).
\end{equation}
The sign ``$+$'' in Eq.~(\ref{eq4}) describes the process of $\hbar
\omega $-quantum absorption (i.e. the energy of charge carriers
increases), and the sign \textquotedblleft $-$\textquotedblright\
corresponds to the emission of a quantum $\hbar \omega $ (i.e. the
energy of charge carriers decreases).

The quantity $P(-)$ describes the variation of the hot hole energy per unit
time, which is related to the emission of a quantum $\hbar\omega$. To
obtain the total energy change induced by the emission of all quanta in a unit
frequency interval into the space angle $d\Omega$, we have to multiply $P(-)$
by the density of final field states in a solid angle $d\Omega$, i.e. by
\begin{equation}
d\rho(\omega)=\frac{V}{(2\pi c)^{3}}\omega^{2}d\Omega.  \label{eq6}
\end{equation}
Then, the product $P(-)d\rho(\omega)$ describes the radiation
emission of charge carriers induced by the electromagnetic wave
field. However, we are interested in the spontaneous radiation
emission by hot carriers. To find this quantity, we use the Einstein
ratio between the probabilities of induced and spontaneous emissions
(see, e.g., work \cite{12}). For this purpose, we must first
normalize the vector-potential ${\bf A}^{(0)}$ (see Eq.~(\ref{eq4}))
in such a way that the volume $V$ would contain $N_{\rm ph}$
photons, i.e. we have to use the condition
\begin{equation}
\label{eq7} \frac{1}{V}N_{\rm ph} \,\hbar \,\omega =\frac{E^2}{4\pi
}=\frac{1}{8\pi }\,\left(\! {\frac{\omega }{c}}\! \right)^{\!2}
A^{(0)2}.
\end{equation}
Whence,
\begin{equation}
\label{eq8} A^{(0)}=2c\left(\! {\frac{2\pi \;\hbar }{V\omega }N_{\rm
ph} }\! \right)^{\!1/2}\!\!.
\end{equation}
After substituting Eq.~(\ref{eq8}) in Eq.~(\ref{eq4}), the quantity
$W^{(-)}\equiv\left[ {P(-)d\rho(\omega)}\right] _{N_{\rm ph}=1}$ is
the power of spontaneous radiation emission by hot carriers in a
unit spectral interval into the solid angle $d\Omega$.

As was shown in work \cite{6}, after the transition in Eq.~(\ref{eq4}) to a
deformed coordinate system, in which the elliptic isoenergetic
surfaces (\ref{eq1}) transform into spherical ones, the integrals in
expression (\ref{eq4}) can be calculated, and we obtain
\[
 W^{(-)}=\frac{e^6n\,N_i \sqrt {m_{\parallel} } }{(2\pi
)^{3/2}c^3}\;\frac{\Psi (\infty )\,d\Omega }{(m_{\parallel} \chi
_{\parallel} -m_\bot \chi _\bot )^2}\times \]\vspace*{-5mm}
\begin{equation}
\label{eq9}
\begin{array}{l}
\times\left\{\! \!{\begin{array}{ll}
 \frac{2\sqrt{\pi}}{\sqrt {\theta _p } }\ln \left(\! {\frac{C_1 ^2\hbar ^2}{8m_\bot \chi
_\bot \theta _p }}\! \right)^{\!-1}& \text{for~}\hbar \omega \ll\theta _p, \\
 \frac{1}{\sqrt {\hbar \omega } }\exp \left(\! {-\frac{\hbar \omega }{\theta
_p }}\! \right)& \text{for~}\hbar \omega \gg\theta _p ,\\
 \end{array}}\!\! \right\}
 \end{array}
\end{equation}
where $\ln C_{1}=0.577\mbox{...}$ is the Euler constant.

According to the results of work \cite{6}, the quantity $\Psi(\infty)$
equals
\[
 \Psi (\infty )=\frac{1}{b_0^3 }\left[ {b_0 +(1-b_0^2 )\,{\rm arctg}\,\frac{1}{b_0
}} \right]\,\sin ^2\varphi + \]
\begin{equation}
\label{eq10}
 + 2\frac{m_\bot }{m_{\parallel} }\left[ {-\frac{1}{1+b_0^2 }+\frac{1}{b_0
}{\rm arctg}\,\frac{1}{b_0 }} \right]\,\cos ^{2\,}\varphi,
 \end{equation}
where $\varphi$ is the angle between the polarization vector and the
axis $C_{3}$ (the latter is parallel to the axis of revolution of
ellipsoids of the effective-mass tensor), and
\begin{equation}
b_{0}=\frac{m_{\bot}\chi_{\bot}}{m_{\parallel}\,\chi_{\parallel}-m_{\bot
}\,\chi_{\bot}}.   \label{eq11}
\end{equation}
At $\chi_{\parallel}=\chi_{\bot}$, the quantity $b_{0}$ coincides with the
corresponding quantity obtained in work \cite{6}.

Substituting expression (\ref{eq10}) in Eq.~(\ref{eq9}), we obtain
\begin{equation}
W^{(-)}=\left\{
{a_{\bot}\,\sin^{2}\varphi+a_{\parallel}{\cos}^{2}\varphi}\right\}
d\Omega.  \label{eq12}
\end{equation}
The coefficients $a_{\bot}$\ and\ $a_{\parallel}$ can be easily obtained by
comparing expressions (\ref{eq9}) and (\ref{eq12}). It is not difficult to
get convinced that if $m_{\bot}=m_{\parallel}$ and
$\chi_{\bot}=\chi_{\parallel}$, we have $a_{\bot}=a_{\parallel},$ and the angular dependence
disappears from expression (\ref{eq12}).

Hence, we have obtained the explicit expression for the angular
dependence of the spontaneous radiation emission by hot holes in the
form (\ref{eq12}). The coefficients that characterize this angular
dependence~-- these are $a_{\bot}$\ and\ $a_{\parallel}$~-- are
known functions of the mass tensor components ($m_{\bot}$ and
$m_{\parallel}$), the dielectric permittivity tensor components
($\chi_{\bot}$ and $\chi_{\parallel}$), the light frequency
$\omega$, and the hot electron temperature $\theta_{p}$. All those
parameters, but $\theta_{p}$, are known.

To obtain the simplest expression for the hot-hole temperature, we confine
the consideration to low temperatures ($4~\mathrm{K}\leq T\leq
20~\mathrm{K}$) and weak enough electric fields, at which $T_{p}\ll 138~\mathrm{K}$, i.e.
when the value of $kT_{p}$ is much lower than the energy of optical phonons.
In those ranges of temperatures and fields, it is possible to consider that
the relaxation of the momentum of holes takes place predominately at their
scattering by ionized impurities, and the energy relaxation occurs at their
quasielastic scattering by acoustic phonons. The energy given by hot
carriers to the lattice per unit time at the quasielastic scattering by
acoustic vibrations, according to the results of work \cite{13}, equals
\begin{equation}
\label{eq13} \int {d\bar {p}\;\varepsilon ({\bf p})\;\hat {I}_{ak}
f=g\,n\,\theta _p^{3/2} \left( \!{1-\frac{\theta }{\theta _p }}\!
\right)}\! ,
\end{equation}
where
 \begin{equation}
g=\frac{8\,\sqrt 2 \;m_\bot  \sqrt {m_{\parallel} } }{\pi ^{3/2}\rho
\,\hbar ^4}\left(\! {\frac{2\,m_\bot ^ +m_{\parallel} }{3}}
\!\right)\sum\nolimits_d^2\, ,
\end{equation}
and $\Sigma _{d}$ is the constant of the deformation potential (we
confine the consideration to the one-constant approximation).

In the stationary case, the scattering power (\ref{eq13}) must be equal
to that obtained by carriers from the heating field ${\bf F}$, i.e.
\begin{equation}
\label{eq14} {\bf j}{\bf F}=e\,n\,\left\{ \!{\;\mu _\bot (\theta _p
)F_\bot ^2 +\mu _{\parallel} (\theta _p )\,F_{\parallel}^2 \,}\!
\right\}\! ,
\end{equation}
where ${\bf j}$ is the current-density vector, $\mu _{\bot }(\theta
_{p})$ and $\mu _{\parallel }(\theta _{p})$ are the transverse and
longitudinal, respectively, components of the mobility tensor
arising owing to the scattering of charge carriers by ionized
impurities \cite{6},
\begin{equation} \mu _{\bot }(\theta
_{p})=\frac{8}{\sqrt{\pi }}\frac{e\,\tau _{\bot }(\theta
_{p})}{m_{\bot }};\quad \mu _{\parallel }(\theta
_{p})=\frac{8}{\sqrt{\pi }}\frac{e\,\tau _{\parallel }(\theta
_{p})}{m_{\parallel }}, \label{eq15}
\end{equation}%
and $\tau _{\bot }(\theta _{p})$ and $\tau _{\parallel }(\theta _{p})$ are
the transverse and longitudinal, respectively, components of the relaxation
tensor at the impurity-driven scattering, given by the expressions
\[
\frac{1}{\tau _\bot (\theta _p )}=\frac{8}{3}\;\frac{e^4\,\sqrt
{2m_{\parallel} } }{\chi _\bot ^2 \;m_\bot \,\theta _p^{3/2} }\times
\]\vspace*{-5mm}
\begin{equation}
\times N_i \frac{b_0 }{2}\left[ {b_0 +(1-b_0^2 )\,{\rm
arctg}\,\frac{1}{b_0 }} \right]\ln \left(\! {\frac{C_1 \,\hbar
^2}{8m_\bot \,\chi _\bot \,\theta _p }}\! \right)^{\!-1}\!,
\end{equation}\vspace*{-5mm}

\noindent and
\[ \frac{1}{\tau _{\parallel} (\theta _p
)}=\frac{8}{3}\frac{e^4\,\sqrt {2m_{\parallel} } }{\chi _\bot ^2
\;m_{\parallel} \,\theta _p^{3/2} }\times\]\vspace*{-5mm}
\begin{equation}
\label{eq16} \times N_i b_0 \left[ {-b_0 +(1+b_0^2 )\,{\rm
arctg}\frac{1}{b_0 }} \right]\;\ln \left(\! {\frac{C_1 \hbar
^2}{8m_\bot \,\chi _\bot \,\theta _p }}\! \right)^{\!-1}\!.
\end{equation}
At $\chi _{\bot }=\chi _{\parallel }$, expressions (18) and
(\ref{eq16}) transform into the corresponding formulas obtained in
work \cite{6} (Unfortunately, the notations $\frac{1}{\tau _{\bot
}}$ and $\frac{1}{\tau _{\parallel }}$ in formula (54) in work
\cite{6} were transposed!)

If the weak dependence on $\theta _{p}$ in the logarithm in formulas
(18) and (19) is neglected, then
Eqs.~(\ref{eq15})--(\ref{eq16}) yield
\begin{equation}
\label{eq17} \mu_{\bot,\parallel } (\theta _p )\approx \left(\!
{\frac{\theta _p }{\theta }}\! \right) ^{\!3/2}\!\mu _{\bot
,\parallel } (\theta ).
\end{equation}
Now, let us make expressions (\ref{eq13}) and (\ref{eq14}) equal to
each other and use relation (\ref{eq17}). Then, from this balance
equation, we easily obtain the temperature of hot holes,
\begin{equation}
\label{eq18} \theta _p =\theta \,\left\{ \!{1-\frac{e}{g\,\theta
^{3/2}}\left[ {\mu _\bot (\theta )\,F_\bot ^2 +\mu _{\parallel}
(\theta )\;F_{\parallel}^2 } \right]}\! \right\}^{\!-1}\!.
\end{equation}
The validity of this formula is restricted to fields, at which
$\theta _{p}<\hbar \omega _{0}$, where $\omega _{0}$ is the
frequency of the lowest optical mode (it corresponds to a
temperature of 138~K). If $\theta _{p}$ approaches the energy of
optical phonons, the scattering of holes by the latter has to be
taken into consideration.

Hence, we obtained an expression for the angular dependence of
the spontaneous radiation emission by hot holes in tellurium in the form
of formula (\ref{eq12}) in the case where the dominating mechanism
of momentum scattering is the scattering by ionized impurities. In so
doing, we made allowance for the tensor character of the effective mass
and the dielectric permittivity in $p$-Te. In work \cite{1}, we obtained
an expression for the spontaneous radiation emission by hot carriers
with the dispersion law (\ref{eq1}) in the case where the acoustic scattering
plays the dominating role. The angular dependence of the spontaneous
radiation emission by hot carriers looks like Eq.~(\ref{eq12}) at
both acoustic- and impurity-driven scattering. The only difference
consists in the different dependences of the coefficients $a_{\bot}$ and
$a_{\parallel}$ on the hot-carrier temperature.

\section{Experiment and Its Discussion}

The experimental setup is shown in Fig.~1. We used specimens of
single-crystalline tellurium grown by the Czochralski method. The
specimens were either cut out of single-crystalline ingots along the
axis $C_{3}$ or carefully sawn out of those ingots across it. Then,
the specimens were annealed in the hydrogen atmosphere at a
temperature of 380$~^{\circ }\mathrm{C}$ for 200~h and treated with
a chromic etching solution
$\mathrm{HF}+\mathrm{CrO}_{3}+\mathrm{H}_{2}\mathrm{O}$ (1:1:3).
After annealing and etching, the mobility and the concentration of
charge carriers fell within the intervals
5200--6000~$\mathrm{cm}^{\mathrm{2}}/\mathrm{V/s}$ and $\left(
1.1\div 1.4\right) \times 10^{15}$~\textrm{cm}$^{-3}$, respectively.
The cross-section dimensions of specimens were $1\times
1.2~\mathrm{mm}^{\mathrm{2}}$, and their length varied from 3.2 to
7~mm. Ohmic contacts were soldered with the use of the solder
50\%~St + 47\%~Bi + 3\%~Sb.

\begin{figure}
\includegraphics[width=5.1cm]{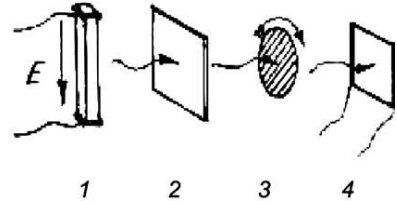}
\vskip-3mm\caption{Experimental setup: $p$-Te specimen with either
parallel ($\Vert $) or normal ($\bot $) $C_{3}$-axis orientation
({\it 1}), black polyethylene filter ({\it 2}), polarizer (analyzer)
({\it 3}), and Ge(Ga) detector ({\it 4})  }\vskip2mm
\end{figure}

To heat up holes in Te specimens, a generator of electric pulses with a low
input resistance of about 20~$\Omega$ was used, which enabled us to carry
out measurements for low-resistance specimens. The pulse duration was
0.8~$\mu\mathrm{s}$ and the repetition frequency was 6~Hz. As a 3-THz radiation
detector ($\lambda\approx100$~$\mu\mathrm{m}$), we used a Ge(Ga) detector
$4\times5\times1.5~\mathrm{mm}^{3}$ in size.

A signal registered by the detector was amplified with the use of a
broadband amplifier, integrated, converted into a constant voltage,
and supplied to a two-coordinate recorder. The short-wave section of
the radiation spectrum emitted by hot holes (\mbox{$\lambda
<50$~$\mu \mathrm{m}$}) was cut off by applying a filter fabricated
of black polyethylene. The polarizer (analyzer, Fig.~1) was rotated
at a low speed of one~rotation per 2~min, and its axis was rigidly
connected with the long-axis direction of the emitting specimen
(and, hence, with the direction of the electric field applied to the
specimen). As the \textquotedblleft zero\textquotedblright\ rotation
angle of a polarizer, its such orientation was selected, when the
direction of polarizer grooves coincided with the direction of
specimen's long axis and the electric field applied to the specimen.
We recall that the polarizer transmits the electromagnetic wave only
in the case where the electric component of the wave is
perpendicular to its grooves.

\begin{figure}
\includegraphics[width=7cm]{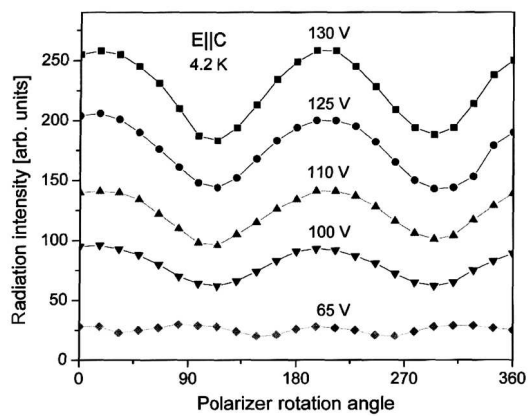}
\vskip-3mm\caption{Dependences of the specimen radiation emission
intensity on the polarizer rotation angle for various heating fields
(indicated in the figure) in the ($p$-Te$\parallel C_{3}$)-geometry
of \mbox{experiment}  }\vskip3mm
\end{figure}

The polarization dependences of the terahertz radiation emission by hot
carriers in $p$-Te were studied using the specimens cut out along
the crystallographic axis $C_{3}$ and perpendicularly to it (below,
we refer to them as the specimens of the first and second types).
Figures~4,$\ a$ and $b$ exhibit one of six (parallel) ellipsoids,
which describe the dispersion law for holes in $p$-Te (\ref{eq1}),
and demonstrate the orientation of the long axis of this ellipsoid
with respect to the crystallographic axis $C_{3}$ and the direction
of electric field applied to the specimen. In both cases, the
electric field is applied along the long axis of the specimen.

\begin{figure}
\includegraphics[width=7cm]{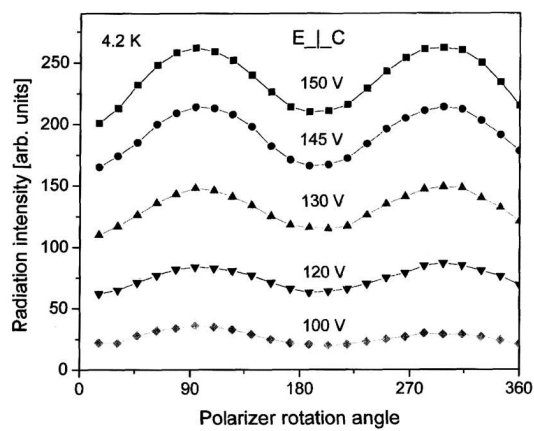}
\vskip-3mm\caption{The same as in Fig.~2, but for the
($p$-Te$\perp$$ C_{3}$)-geometry of experiment  }
\end{figure}

While studying the polarization dependence of the radiation emission by
specimens of both the first and second types, the polarizer was rotated in
the plane $zy$ around the axis $x$. In Figs.~2 and 3, the dependences of
the radiation intensity emitted by hot holes on the angle between the direction
of polarizer grooves and the direction of electric field that heats up
charge carriers are depicted. It will be recalled that the zero value of
this angle corresponds to the situation where the polarizer grooves are
parallel to the applied electric field.

\begin{figure}
\includegraphics[width=4.5cm]{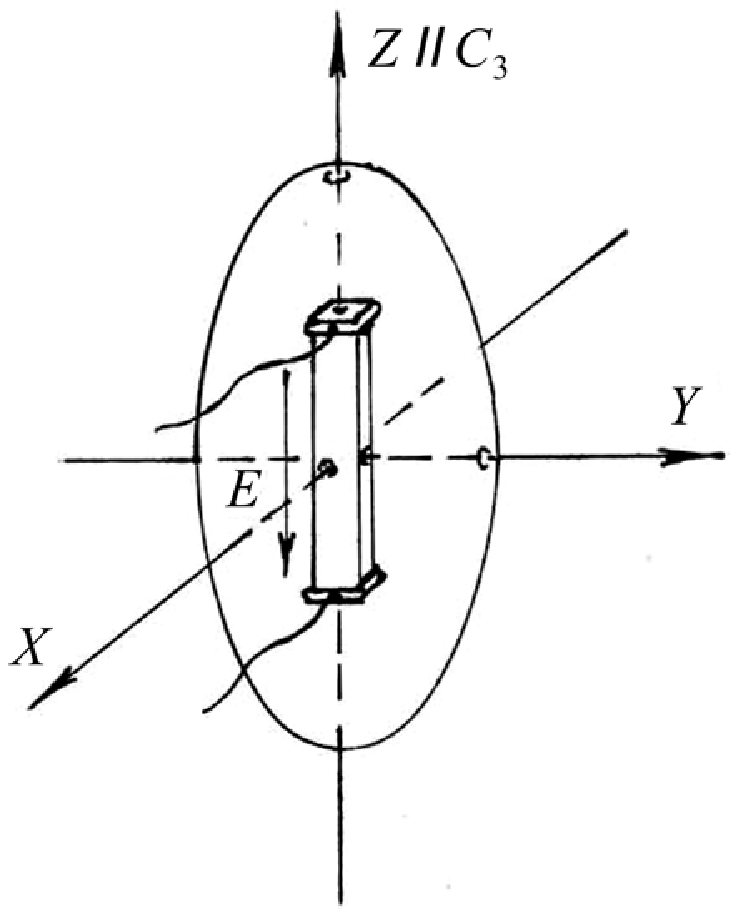}\\
{\large\it a}\\[2mm]
\includegraphics[width=4.7cm]{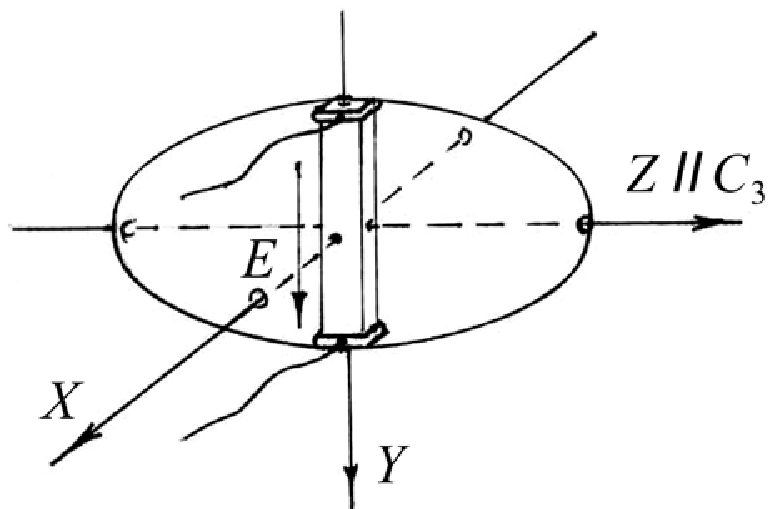}\\
{\large\it b}
 \vskip-3mm\caption{Specimens and their orientation
with respect to the crystallographic axis: specimen cut out ({\it
a})~along the axis and ({\it b}) perpendicularly to it  }
\end{figure}

In the theory developed above, the angular dependence of the radiation intensity
was described by the angle between the axis $C_{3}$, which is parallel to
the long axis of the ellipsoid, and the polarization direction, i.e. the
direction of the electric component of an emitted electromagnetic wave. Since the
polarizer transmits only the electric wave component which is perpendicular
to its grooves, then, as is seen from Fig.~4,$a$ corresponding to the case
of first-type specimens, when the heating electric field, the axis $C_{3}$,
and the grooves are parallel to one another, the angles in Fig.~2 and in
the theoretical formula (\ref{eq12}) are shifted with respect to one another by
$\pi/2$. At the same time, for specimens of the second type, for which
the heating field and the axis $C_{3}$ are mutually perpendicular, the
angles in Fig.~3 and in the theory (formula~(\ref{eq12})) coincide (see
Fig.~4,$b$). In this case, there is a minimum at the angle $\varphi=0$. This
fact agrees with formula (\ref{eq12}). Passing in this formula to the
doubled angle, we obtain
\begin{equation}
W^{(-)}=\frac{1}{2}\{a_{\bot}+a_{\parallel}+(a_{\parallel}^{-}a_{\bot})\,%
\cos\,2\varphi\,\}d\Omega.   \label{eq19}
\end{equation}
Whence, one can see that the presence of the minimum at $\varphi=0$
corresponds to the condition \mbox{$a_{\bot}/a_{\parallel}>1$}.

From Eqs.~(\ref{eq9}) and (\ref{eq10}), we obtain
\begin{equation}
\label{eq20} \frac{a_\bot }{a_{\parallel} }=\frac{\frac{1}{b_0^3
}\left[ {b_0 +(1-b_0^2 )\,{\rm arctg}\,\frac{1}{b_0 }}
\right]}{2\frac{m_\bot }{m_{\parallel} }\,\left[ {-\frac{1}{1+b_0^2
}+\frac{1}{b_0 }\,{\rm arctg}\,\frac{1}{b_0 }} \right]}\,.
\end{equation}
For $p$-Te, $m_{\parallel}=0.26\,m_{0}$,
$m_{\bot}=0.11\,m_{0}$,$\;\chi _{\parallel}=56$, and
$\chi_{\bot}=33$. Hence, $b_{0}^{2}\equiv\frac{m_{\bot
}\chi_{\bot}}{m_{\parallel}\chi_{\parallel}-m_{\bot}\chi_{\bot}}\approx0.35$
and $\frac{a_{\bot}}{a_{\parallel}}\approx7$. Thus, in both the
theory and the experiment, we obtain a minimum at $\varphi=0$.
Therefore, the positions of minima and maxima in the theory and the
experiment coincide. Hence, our theory adequately predicts a
periodic character of polarization dependences of the spontaneous
radiation emission by hot holes in $p$-Te and correctly evaluates
the positions of maxima and minima for this radiation.

It would be of interest to have not only a qualitative, but also a
quantitative comparison between the theory and the experiment. However,
unfortunately, it cannot be done now. First, expression (\ref{eq12}) gives
the spectral distribution of the radiation intensity, whereas experimentally the
integrated radiation power in the given frequency range is measured.
Therefore, for the quantitative comparison to be made, expression
(\ref{eq9}) should be integrated over a frequency interval given in the experiment.
This procedure renormalizes the coefficients $a_{\bot}$ and $a_{\parallel}$
in expression (\ref{eq12}). While integrating over the frequency, we would
have to determine the temperature of hot carriers, $\theta_{p}$, for every
heating field, which is a separate large problem.

Second, the very expression for $\theta_{p}$ can change with the growth of
the electric field owing to the engaging of new relaxation mechanisms. However,
the angular dependence of the spontaneous radiation emission by hot carriers in
form (\ref{eq12}) has a more universal character, which is governed, first
of all, by the dispersion law (\ref{eq1}). Provide that the dispersion law is
fixed, only the specific values of parameters $a_{\bot}$ and $a_{\parallel}$
in formula (\ref{eq12}) depend on the scattering mechanism.

\section{Conclusions}

The dependences of the terahertz radiation emission intensity by hot
charge carriers in $p$-Te on the angle between the crystallographic axis
$C_{3}$ and the polarization vector, as well as their periodic character,
have been studied both theoretically and experimentally. The periodic
character of the dependence concerned and the positions of radiation
intensity minima and maxima were demonstrated to coincide in the theory and
the experiment. The polarization dependences of the radiation emission by hot
charge carriers were found to be associated with the anisotropy of the charge
carrier dispersion law and the anisotropy of the dielectric permittivity.

\vspace*{-3mm}
\rezume{%
П.М. Томчук, В.М. Бондар, Л.С. Солончук}{ПОЛЯРИЗАЦІЙНІ ЗАЛЕЖНОСТІ
ТЕРАГЕРЦОВОГО\\ ВИПРОМІНЮВАННЯ ГАРЯЧИМИ НОСІЯМИ \\ЗАРЯДУ в $p$-Te}
{В роботі теоретично і експериментально вивчено поляризаційні
залежності терагерцового випромінювання гарячими носіями заряду в
$p$-Te. Показано, що кутові залежності спонтанного випромінювання
гарячих носіїв зумовлені анізотропією їх закону дисперсії і
анізотропією діелектричної проникності. Встановлено, що
поляризаційні залежності випромінювання визначаються кутом між
кристалографічною віссю $C_3 $ в $p$-Te та ортом поляризації і ці
залежності мають періодичний характер.}


\begin{thebibliography}{99}
\bibitem{1} V.M.~Bondar, O.G.~Sarbei, and P.M.~Tomchuk, Fiz. Tverd. Tela
\textbf{44}, 1540 (2002).

\bibitem{2} V.M.~Bondar and N.F.~Chernomorets, Ukr. Fiz. Zh. \textbf{48}, 51
(2003).

\bibitem{3} P.M.~Tomchuk and V.M.~Bondar, Ukr. Fiz. Zh. \textbf{53}, 668
(2008).

\bibitem{4} T.H.~Mendum and R.N.~Dexter, Bull. Amer. Phys. Soc. \textbf{9},
632 (1961).

\bibitem{5} R.V.~Parfen'ev, A.M.~Pogarskii, I.I.~Farbshtein, and
S.S.~Shalyt, Fiz. Tverd. Tela \textbf{4}, 3596 (1962).

\bibitem{6} P.M.~Tomchuk, Ukr. Fiz. Zh. \textbf{49}, 682 (2004).

\bibitem{7} M.S.~Bresler, V.G.~Veselago, Yu.V.~Kosichkin, G.E.~Pi\-kus,
I.I.~Farbshtein, and S.S.~Shalyt, Zh. Eksp. Teor. Fiz. \textbf{57}, 1479
(1969).

\bibitem{8} M.S.~Bresler and D.S. Mashovets, Phys. Status Solidi~B
\textbf{39}, 421 (1970).

\bibitem{9} A.S.~Dubinskaya and I.I.~Farbshtein, Fiz. Tverd. Tela
\textbf{8}, 1884 (1966).

\bibitem{10} P.M.~Gorlei, P.M.~Tomchuk, and V.A.~Shenderovskyi, Ukr. Fiz.
Zh. \textbf{20}, 705 (1975).

\bibitem{11} P.M.~Gorlei, V.S.~Radchenko, and V.A.~Shen\-de\-rov\-skyi,
\textit{Transport Processes in Tellurium} (Nau\-ko\-va Dum\-ka,
Kyiv, 1987) (in Russian).

\bibitem{12} A.S.~Davydov, \textit{Quantum Mechanics} (Pergamon Press, New
York, 1976).

\bibitem{13} I.M.~Dykman and P.M.~Tomchuk, \textit{Transport Phenomena and
Fluctuations in Semiconductors} (Naukova Dumka, Kyiv, 1981) (in
Russian).\vspace*{-2mm}

\begin{flushright}
{\footnotesize Received 27.06.12.\\ Translated from Ukrainian by
O.I.~Voitenko}
\end{flushright}
\end{thebibliography}
\end{document}